\documentclass[aps,prd,amsmath,amssymb,twocolumn,nofootinbib,showpacs,preprintnumbers]{revtex4-1}
\usepackage{graphics,hyperref,color,bbold,multirow}
\newcommand{\etal}{{\it et al.}}
\hypersetup{ 
    pdfnewwindow=true,   
    colorlinks=true,     
    linkcolor=blue,      
    citecolor=blue,      
    filecolor=blue,      
    urlcolor=blue        
}  

\begin{document}

\title{Charmonium spectrum with a Dirac potential model in the momentum space} 

\author{David Molina}
\email{djmolinab@unal.edu.co}
\affiliation{Universidad Nacional de Colombia, Bogot\'a 111321, Colombia}
\author{Maurizio De Sanctis}
\affiliation{Universidad Nacional de Colombia, Bogot\'a 111321, Colombia}
\author{C\'esar Fern\'andez-Ram\'{\i}rez}
\email{cesar.fernandez@nucleares.unam.mx}
\affiliation{Instituto de Ciencias Nucleares, Universidad Nacional Aut\'onoma de M\'exico, 
Ciudad de M\'exico 04510, Mexico}

\begin{abstract}
We study the charmonium spectrum using a complete 
one gluon exchange approach based 
on a phenomenological relativistic $q\bar{q}$ 
potential model with Dirac spinors in momentum space.
We use phenomenological screening factors to include 
vacuum quantum effects. 
Our formulation does not rely on nonrelativistic approximations. 
We fit the lowest-lying charmonia (below the $D\bar{D}$ threshold) and predict the 
higher-lying resonances of the spectrum. 
In general, we reproduce the overall structure of the charmonium spectrum and,
in particular, we can reasonably describe the $X(3872)$ resonance mass as (mostly) a
$c\bar{c}$ state. 
The numerical values of the free parameters of the model are determined 
taking into account also
the  experimental uncertainties of the resonance energies.
In this way, we are able to obtain the uncertainties of the theoretical resonance masses 
and  the correlation among the free parameters of the model.
\end{abstract}
\date{\today}
\maketitle

\section{Introduction} \label{sec:introduction}
In the last years the  interest in charmonia
has been renewed  by the experimental observation of new  charmoniumlike 
states~\cite{Olsen1,brambilla,Godfrey:2008nc,Esposito:2016noz,Swanson1,Esposito:2014rxa,Lebed}. 
Some of these states have been easily identified as charmonia, 
e.g. $h_c(1P)$~\cite{Rosner,Rubin} and $\chi_{c2}(2P)$~\cite{Uehara}, 
as reported by the Particle Data Group (PDG)~\cite{PDG2016}.
However there exist other  important resonances that are still under examination.
These states, denoted as $X$, $Y$ and $Z$,  do not match the predictions of the  nonrelativistic 
or semirelativistic $q\bar{q}$ potential models 
and some of them have been described as  tetraquarks, molecules, glueballs and 
hybrids~\cite{Olsen1,brambilla,Esposito:2016itg,Brodsky:2014xia,Godfrey:2008nc,Pilloni:2015doa,Esposito:2016noz,Lebed,Guo:2008yz,Guo-Galata,Ber-Brambi,Faccini:2013lda}.
Another possible explanation  suggests that some of these
states may be the result of kinematic effects~\cite{Szczepaniak:2015eza,Guo:2014iya}
that appear in the nearby of the thresholds of the open charm channels. 
The interpretation of these resonances depends, to a certain  extent, 
on the theoretical approach that is followed to perform the amplitude 
analysis of the data~\cite{Pilloni:2016obd,Kang:2016jxw}
and to study charmonium spectroscopy.

Quantum chromodynamics (QCD)  must be considered, in any case, the underlying field theory 
for the study of these systems.
However, due to the difficulty of finding analytically (approximated) solutions 
for QCD, alternative approaches have been followed.
For example, numerical lattice QCD (LQCD)~\cite{Liu:2012ze,Christopher,Lang,Gunnar},
potential quark 
models~\cite{Isgur,Capstick,Faustov,DeSanctis2,Cao2012,DeSanctis3,klink,poly,Bram-Gro,Kawanai1,Laschka,PhysRevD.85.091503,Kawanai2},
unquenched quark models (UQM)~\cite{Santo,Santo2,Ferre-Galata,FeSa1,FeSa2,FeSa3},
and Bethe-Salpeter approaches~\cite{Hilger:2015hka,Fischer,Popovici}.

LQCD represents 
a very promising but, at the same time, highly difficult tool.
LQCD computations reproduce the structure of the
charmonium spectrum and predict many undiscovered
states~\cite{Liu:2012ze,Christopher,Gunnar}.
In particular, 
LQCD is computationally 
taxing, especially when open channels and 
the resonant nature of the states are considered,
and still needs to be improved in order to reproduce accurately
the  hadron 
spectrum~\cite{Luscher,Rummukainen,Briceno1,Briceno2,Dudek:2012xn,mesonsector,Wilson,Dudek}. 
Exploratory LQCD simulations of the $D\bar{D}$
scattering in S and P waves 
have been recently computed~\cite{Lang} but they
still need to be improved to achieve the precision
of the lowest-energy meson sector~\cite{Briceno:2016mjc}.
However, the construction of potential quark models
whose interaction has been derived from 
LQCD~\cite{Kawanai1,Laschka,PhysRevD.85.091503,Kawanai2},
building up models that are not purely
phenomenological~\cite{Bali:2000gf},
has proved to be
a very useful tool for the study of these physical systems.
Historically, the Cornell potential~\cite{Eichen} represents 
the first  interaction  that was introduced, 
with partial success, to study the charmonium spectrum.
This model consists of a Coulomb-like 
potential and a linear term inserted in a 
standard nonrelativistic Schr\"odinger equation. 
The Coulomblike  term takes into account reduction of the 
one gluon exchange (OGE) in the short range region.
The linear term introduces quark confinement in the model, 
in approximate agreement with the  
spectroscopy of the first excited states.
LQCD studies have confirmed the presence of the linear term 
in the long range effective interaction~\cite{PhysRevD.85.091503}.
However, in this model, the nonrelativistic approximation cannot be really justified 
and the fine structure of the spectrum,
related to the spin-spin, spin-orbit and tensor interactions, has to 
be introduced by hand in a purely phenomenological way.
Furthermore, higher order effects of the field theory reduction are also absent. 

Throughout the literature  several \textit{relativized}
quark models have been introduced  in order to 
incorporate some relativistic
effects~\cite{Isgur,Capstick,Faustov,DeSanctis2,Cao2012,DeSanctis3,klink,poly,Bram-Gro}.
A possible approach to include some of the relativistic effects 
in a consistent way is given by the generalization of the 
Fermi-Breit expansion 
(originally studied for the electromagnetic interaction) 
followed in~\cite{DeSanctis2,Cao2012,DeSanctis3}
to study the resonant states of strongly interacting quarks.
We point out that this approach is not fully relativistic because
the interaction terms are expanded in powers of $p/m$, where  
$p$ and $m$ are the quark momentum and mass, respectively. 
These models also assume that the mean values of $p/m$ are small and, consequently,  
only the first terms of the expansion are retained.
The obtained effective interaction, by means of a Fourier transform, 
can be expressed in the coordinate space in terms of local operators.
As a consequence, the eigenvalue equation for the effective Hamiltonian of these models 
can be solved with the standard techniques.
For the specific case of relativized quark models for hadronic systems,
the Bakamjian-Thomas construction may be used in the context of
relativistic Hamiltonian dynamics~\cite{klink,poly}. 
This approach allows us 
to perform exact Lorentz boosts on the wave functions of the model,
but no connection between the interaction Hamiltonian and the Dirac structure of the underlying theory
can be easily established.
In other words, one cannot express 
the interaction of the model in terms of the 
Dirac covariants that are the building blocks of 
any relativistic theory, 
i.e. the scalar, vector, 
pseudoscalar, tensor and pseudotensor terms.
Further details can be found in~\cite{klink,poly}.

Another approach to improve the dynamical description of the 
quarkonium spectrum consists in employing the
UQM formalism.
In the UQM, starting from a relativized quark model, the coupling of the valence quarks
to the continuum is taken into account through a self energy term~\cite{Santo,Santo2,Ferre-Galata}. 
This coupling causes a shift in the energy, generally leading to a reduction in the masses 
of the quarkonia~\cite{FeSa1,FeSa2,FeSa3}
in the high-lying states.
This effect can also be incorporated, in a more phenomenological way, 
through the screening of the interaction potential~\cite{Li,Gonzalez} 
which produces a similar global result.

It is  also possible to perform a four-dimensional fully relativistic computation of the 
heavy meson spectra solving the Bethe-Salpeter equation.
This approach has been taken in the context of the \textit{rainbow ladder} 
approximation~\cite{Hilger:2015hka}.
This approach presents relevant theoretical advantages; 
however, the quality of the theoretical spectrum is not completely satisfactory
and, in particular, spurious states are obtained~\cite{Fischer,Popovici}
as a consequence of the adopted dynamical approximations~\cite{Ahlig,Bijtebier1}.

In the present work we develop a relativistic model inspired by the OGE interaction. 
In our calculation we do not perform any nonrelativistic expansion.
Hence, the well-known spin-orbit, spin-spin and tensor effects are automatically given by the
relativistic calculation.
Due to the Dirac structure of the interaction, the model is necessarily constructed in the
momentum space.
We include vector and scalar interaction terms as well as 
phenomenological screening factors~\cite{Li,Gonzalez,Bai}.
We solve an integral eigenvalue Hamiltonian equation in the momentum space through
a variational method and we perform the diagonalization of the 
Hamiltonian matrix in a selected orthonormal basis.
We determine the parameters of the model by fitting the 
low-lying charmonia states taking into account the experimental uncertainties.
These uncertainties are exactly propagated to the parameters 
and to the computed spectrum using the bootstrap technique~\cite{NumericalRecipes,Landay:2016cjw}.
This method is computationally expensive but allows a rigorous determination, 
from the statistical point of view, of the parameters of the model and their uncertainties,
as well as the propagation of such uncertainties to the computation of the spectrum.
The rigorous study of the uncertainties allows to identify the statistically
significant differences between the theoretical and the experimental spectra,
determining which states are in need of physics beyond the $c\bar{c}$ picture to be described.
Moreover, the method allows to study the correlations among the parameters of the model,
which provides insight on how physically independent are the different 
pieces that build the phenomenological model. 
As far as we know, no previous works on phenomenological models of charmonia have addressed
the problem of rigorously establishing the correlations among parameters,
the associated uncertainties of the parameters and their impact in the predicted spectrum.

The paper is organized as follows. In Section~\ref{sec:model} 
we develop the general structure of our theoretical model.
In particular, we show the details of the vector and scalar interactions that 
we use to obtain the charmonium spectrum. 
In Section~\ref{sec:determspectr} we discuss the methods used to solve 
the eigenvalue equation  and to determine the free parameters of the model and their uncertainties. 
We also show how these uncertainties are propagated 
to the energy levels of the theoretical charmonium spectrum.
In Section~\ref{sec:charm} we provide our results compared to the available experimental data 
and we discuss  different phenomenological interpretations.
Finally, in Section~\ref{sec:conclusions} we summarize the results of the study.
The details on the spin-angle matrix elements are left for Appendix~\ref{appendix}.

\section{Theoretical model}\label{sec:model}
In this section we provide a detailed description of our theoretical model for the study
 of charmonium spectroscopy.
The model consists in a three-dimensional reduction 
of the Bethe-Salpeter equation in which only the contributions
of the positive energy  Dirac spinors of the quarks are retained.
Dirac spinors guarantee that relativity is satisfied and the Dirac structure
of the interaction is highlighted.
Due to the relativistic nature of the model,  the well-known  spin-orbit, spin-spin and tensor effects
are all automatically included in our calculation.
If we expand our model in powers of $p/m$ we recover the standard 
semirelativistic Fermi-Breit interaction~\cite{Lifshitz}.

\subsection{Definitions}
As it is customary, we work in the center of mass (CM) reference frame
of the $c\bar{c}$ system and we use the system of natural units ($\hbar=c=1$).
The quantities of the ket and bra states are labeled with
the indices $a$ and $b$, respectively. Hence, the three-momentum  of the quark (particle 1) 
and the three-momentum of the antiquark (particle 2) are written as 
\begin{equation}\label{defsmom}
\begin{split}
\vec{p}_{1a} &= - \vec{p}_{2a}=\vec{p}_a ,\\
\vec{p}_{1b} &= - \vec{p}_{2b}=\vec{p}_b,
\end{split}
\end{equation}
where the relative three-momenta $\vec{p}_b$ and $\vec{p}_a$ are the actual 
kinematic variables employed in the calculations.
We also introduce the three-momentum transfer
\begin{equation}\label{defq}
\vec q= \vec{p}_b- \vec{p}_a ,
\end{equation}
and the
cosine  of the angle between $\vec{p}_a$ and $\vec{p}_b$
\begin{equation}\label{cosx}
x=\cos\theta=\hat{p}_a\cdot\hat{p}_b, 
\end{equation}
so that
\begin{equation}\label{eq:q32}
\vec{q}^{\: 2}= \vec{p}_a^{\: 2} + \vec{p}_b^{\: 2} -2 p_a p_b x ,
\end{equation}
where $p_a$ and $p_b$ are the moduli of the respective three-momenta,
$\hat{p}_a=\vec{p}_a\slash p_a$ and $\hat{p}_b=\vec{p}_b\slash p_b$.
The on shell  energies of both particles ($i=1, 2$) are
\begin{equation}\label{defe}
\begin{split}
E(\vec{p}_{i a})= E(\vec p_a)=\sqrt{ \vec{p}_a^{\: 2} +m^2}, \\
E(\vec{p}_{i b})= E(\vec p_b)=\sqrt{ \vec{p}_b^{\: 2} +m^2},
\end{split}
 \end{equation}
where $m$ stands for the quark and antiquark masses.
We introduce the energy difference (for both particles) between the bra and ket states
\begin{equation}\label{delte}
\Delta E= E(\vec p_b) -E(\vec p_a),
\end{equation}
and the invariant four-momentum transfer squared ($Q^2>0$)
\begin{equation}\label{defq2}
Q^2=\vec{q}^{\: 2}-(\Delta E)^2.
\end{equation}

Finally, for completeness, we give the standard definition of the positive energy Dirac spinors
that will be used in the calculation
\begin{eqnarray}\label{defdspin}
u(\vec{p}_{i },\vec{\sigma}_i)=\sqrt{\frac{E(\vec{p}_i)+m}{2 E(\vec{p}_i)}}
\begin{pmatrix}
1\\
\frac{\vec{p}_i   \cdot\vec{\sigma}_i }{E(\vec{p}_i)+m}
\end{pmatrix}
\chi,
\end{eqnarray}
where $i=1,2$ is the particle index, $\vec \sigma_i$ are the Pauli matrices
and $\chi$ represents a generic two component spin wave function.
Using Eq.~(\ref{defsmom}), we take $\vec p_i= \vec{p}_{ia}$ and $\vec{p}_i= \vec{p}_{ib}$, 
for the ket and bra states, respectively.
The spinors are normalized to ${u}^{\dag} (\vec{p}_{i },\vec{\sigma}_i)u(\vec{p}_{i },\vec{\sigma}_i) =1$.

\subsection{Total interaction Hamiltonian and screening factors}\label{TotHamil}
The matrix elements of the total Hamiltonian in momentum space
are obtained adding two terms that represent the vector and the scalar interactions
\begin{equation}\label{hint}
\bar {\cal H}_\text{int} (\vec p_b, \vec p_a)= \langle\vec{p_b}|H^{(v)}|\vec{p_a}\rangle +\langle\vec{p_b}|H^{(s)}|\vec{p_a}\rangle, 
\end{equation}
where, for brevity, the spin matrices have not been explicitly written.
However, the standard phenomenological potentials
do not allow for a good  reproduction of  the spectrum,
in particular above the open charm threshold~\cite{Li,Vija-Gon-Gar}.
A more complete dynamical treatment is necessary.
Some of the vacuum quantum effects predicted by QCD
can be taken into account by \textit{unquenching} 
the quark model as in~\cite{bijker,ref,Ferretti:2015fba}. 
In particular,  the virtual pair creation effects are incorporated
to the Hamiltonian, i.e, the contribution given by the coupling of the $c\bar{c}$ states to the 
meson-meson continuum~\cite{Santo,Santo2}.  
Below threshold, this coupling to the continuum states gives rise to virtual $q \bar{q}-q \bar{q}$ components 
in the meson wave function that shift the charmonia~\cite{Santo2} 
and bottomonia~\cite{FeSa1,FeSa2,FeSa3} masses
through a self-energy term.
Part of these  virtual pair production effects can be  taken into account, 
in a more phenomenological way,
introducing a  screening factor in the confinement interaction 
in the coordinate space~\cite{Li,Gonzalez}.

In this work, we introduce this screening effect phenomenologically,
incorporating momentum dependent factors in the interaction Hamiltonian as follows
\begin{equation}
\bar {\cal H}_\text{int} (\vec p_b, \vec p_a)  \rightarrow {\cal H}_\text{int} (\vec p_b, \vec p_a)
=  F_s(p_b) \bar {\cal H}_\text{int} (\vec p_b, \vec p_a)   F_s(p_a), \label{25}
\end{equation}
where the factors $F_s(p_a)$ and $F_s(p_b)$ take into account  the screening effect.
In the present work we use the following stepwise function to describe such screening effects
\begin{equation}
F_s(p)=\frac{1+k_s}{k_s+ \exp{ \left( p^2/p^2_s \right)}}, \label{26}
\end{equation} 
where  $k_s$ and $p_s$ will be determined (as the rest of the parameters of the model) 
by fitting to the lowest energy levels of the charmonium spectrum.
The details on the fit are provided in Section~\ref{sec:fits}.

\subsection{Vector interaction}
In the Coulomb gauge, the vector interaction matrix element reads
\begin{equation}
\langle \vec p_b|H^{(v)}|\vec p_a\rangle =
\alpha  \left( J^0_1 \:J^0_2\: D_{00}+
 J^\alpha_1\: J^\beta_2 \: D_{\alpha \beta}  \right). \label{eq1}
\end{equation}
If we choose $\alpha=\alpha_{em} $ (the electromagnetic fine structure constant) 
we obtain the standard electromagnetic interaction whose expansion up terms $p/m$ 
gives the Fermi-Breit interaction~\cite{Lifshitz}.
We generalize the previous equation to the strong interaction between a quark 
and an antiquark~\cite{DeSanctis2}. In particular,
$J^\mu_i$ represents the quark ($i=1$)
and antiquark ($i=2$) Dirac four-current that will be given explicitly in Eq.~(\ref{fourcurr}). 
Furthermore
\begin{alignat}{2}
D_{00}&=-\frac{1}{\vec{q}^{\: 2}} \label{eq2}, \\
D_{\alpha \beta}&=\frac{1}{Q^2}\left(\delta_{\alpha \beta}-\frac{q_\alpha q_\beta}{\vec{q}^{\: 2}}\right).
\label{eq2b}
\end{alignat}

Given that we use positive energy Dirac spinors,
we can write the continuity equation in the form
\begin{equation}
(\Delta E)\: J^0_{1,2}=\pm\: \vec{q}\cdot\vec{J}_{1,2} , \label{eq3}
\end{equation}
where the signs $+$ and $-$ correspond to the quark and antiquark cases respectively.
Using Eq.~(\ref{eq3}) we can rewrite Eq.~(\ref{eq1}) for the strong interaction as
\begin{equation}
\begin{split}
\langle\vec{p_b}|H^{(v)}|\vec{p_a}\rangle =& \:
V^{(v)}(\vec q)\: \left[ J^0_1\: J^0_2 \left(1-\frac{(\Delta E)^2}{Q^2} \right)\right. \\
&-\left. \vec{J}_1\cdot\vec{J}_2\left(1+\frac{(\Delta E)^2}{Q^2} \right)\right] .
\end{split} \label{eq4}
\end{equation} 

The explicit form of
the effective vector potential $V^{(v)}(\vec q)$
is provided in Eq.~(\ref{vector}) of  Section~\ref{potentials}.
Notice that, compared to Eq.~(\ref{eq1}), 
we take $\alpha=\alpha_{st}$,
which represents the phenomenological strong coupling constant of the model. 

Finally, the quark and antiquark four-currents are given by the standard form
\begin{equation}\label{fourcurr}
J_i^{\mu}=  J_i^{\mu} (\vec \sigma_i; \vec p_b,\vec p_a)
=\bar{u} (\vec{p}_{i b},\vec{\sigma}_i)  \gamma_i^{\mu}u(\vec{p}_{i a },\vec{\sigma}_i) .
\end{equation}

In this work we do not include a chromomagnetic current term, which seems to be unnecessary given
the quality of the obtained spectrum as shown in Section~\ref{sec:charm}.

\subsection{Scalar interaction}
In hadron spectroscopy 
the spin-obit effect is in general rather small
and cannot be reproduced if we only use a vector interaction.
Hence, a  scalar interaction,
partially responsible 
for quark confinement~\cite{Swanson1,Swanson2},
is included in the interaction Hamiltonian 
to improve the description of the
charmonium spectrum~\cite{Cao2012,Ebert}.
Also in the present work 
we find that the scalar interaction is needed to
obtain a good description of the charmonium spectrum. 
We write the scalar interaction in the following standard way
\begin{equation}
\langle \vec p_b|H^{(s)}|\vec p_a\rangle=V^{(s)}(\vec q) \: I_{1} \: I_{2} ,
\label{scalint}
\end{equation}
with the explicit form of the effective scalar potential  $ V^{(s)}(\vec q)$
provided in  Eq.~(\ref{scalar}) of  Section~\ref{potentials}. 
The scalar vertex is given by
\begin{eqnarray}\label{scalvert}
I_i=I_i (\vec \sigma_i; \vec p_b,\vec p_a)=\bar{u} (\vec{p}_{i b},
\vec{\sigma}_i)  u(\vec{p}_{i a },\vec{\sigma}_i).
\end{eqnarray}

\subsection{Vector and scalar potentials}\label{potentials}
In order to include different physical effects in our effective interaction, it is necessary to use  
phenomenological expressions for the  potentials in momentum space. 
There exist a wide range of models for the scalar 
and vector potentials~\cite{DeSanctis2,Faustov,Radford}. 
In this work we use the following effective potentials:
\begin{subequations}\label{vtotq}
\begin{equation}\label{vector}
V^{(v)}(\vec q)=-\frac{4}{3}\frac{\alpha_{st}}{\vec{q}^{\: 2}}
+\beta_{v}\frac{3b^2-\vec{q}^{\: 2}}{(\vec{q}^{\: 2}+b^2)^3} ,
\end{equation}
\begin{equation}\label{scalar}
V^{(s)}(\vec q)=A +\beta_{s} \frac{3b^2-\vec{q}^{\: 2}}{(\vec{q}^{\: 2}+b^2)^3} .
\end{equation}
\end{subequations}
We note that only by a combination of scalar and vector
confining potentials is it possible to eliminate the empirically
contraindicated spin-orbit interaction in the interquark potential.
The vector interaction, Eq.~(\ref{vector}), 
is given by a regularized Cornell potential in momentum space
where the $-\frac{4}{3} \alpha_{st} \slash \vec{q}^{\: 2}$ term in Eq.~(\ref{vector}) corresponds to the
standard pure Coulombian strong potential, being  $\alpha_{st}$ the effective strong coupling constant
and $\frac{4} {3} $ the color factor.
The second term corresponds to  a confining interaction where the  $\beta_v$ parameter 
is related to the vector confinement intensity.
The parameter $b$ has been introduced
to avoid the  $1/{ q}^4 $ singularity in the momentum space interaction
and, in the present model, has no direct physical meaning.
The expression of the vector potential in the coordinate space reads
\begin{equation}\label{vvr}
 V^{(v)}(r)=-\frac{4}{3}\frac{\alpha_{st}}{r}+\beta_{v}\pi^2  r  \exp{(-r b)} ,
\end{equation}
where it is apparent that the second term represents  a linear confinement interaction 
regularized with a decreasing exponential function. 
For convenience, we fix $b=10^{-2}~ \text{GeV}$, 
that corresponds to $1/b\simeq 20~\text{fm}$.
The $b$ value is chosen small enough to avoid
numerical problems and large enough to approximately recover the
Cornell potential for the energy region of interest.
Regarding the scalar interaction in Eq.~(\ref{scalar}), we 
take a phenomenological constant  term, represented by $A$,  
plus  a linear regularized potential, 
analogous to the confining term of the vector potential. 
We note that $\beta_s$ is (as $\beta_v$) related to the confinement strength, 
and we also fix $b=10^{-2}~ \text{GeV}$.
In this work we use two different prescriptions for the scalar potentials 
to fit the experimental data.
In the first prescription (potential I), we set
$\beta_s=0$, 
i.e. we keep the effective constant scalar confinement term $A$ in Eq.~(\ref{scalar})
and we omit the scalar linear regularized potential.
In the second one (potential II), the value of  $\beta_s \neq 0$
is determined fitting the data.
We study these two prescriptions separately in order 
to assess the impact of the confinement term of the scalar interaction.
In what follows we refer to these two models as
\begin{equation}\label{pot_options}
\begin{cases}  
\text{potential I\phantom{I}} \rightarrow \text{model using Eqs.~(\ref{vtotq}) with } \beta_s=0 , \\
\text{potential II} \rightarrow  \text{model using Eqs.~(\ref{vtotq}) with } \beta_s\neq0 .
\end{cases}
\end{equation}

\subsection{Integral equation for the $c \bar c$ system}
Now, we can write the total integral Hamiltonian equation (in momentum space) for our system.
First, we introduce the standard kinetic energy of the quark and the antiquark in the charmonium CM
\begin{equation}\label{kine}
 K(\vec p)= 2\sqrt{\vec{p}^{\: 2} + m^2}  .
\end{equation}

The integral Hamiltonian equation takes the form
\begin{equation}\label{eigen2}
[K(\vec{p_b})+M_0]\Psi(\vec{p}_b)+\int d^3p_a  {\cal H}_\text{int} (\vec{p}_b,\vec{p}_a)\Psi(\vec{p}_a)=M\: \Psi(\vec{p}_b) ,
\end{equation}
where ${\cal H}_\text{int} (\vec{p}_b,\vec{p}_a)$ is given by Eq.~(\ref{hint}), 
$M_0$ represents a phenomenological zero point energy of the spectrum, $M$ is the resonance mass 
(that is the eigenvalue of the integral equation) and $\Psi(\vec{p})$ is the resonance wave function.
For brevity, in the interaction Hamiltonian, 
the Pauli spin matrices have not been explicitly written and also, in the wave function,
the energy and angular momentum quantum numbers have been omitted.

Equation~(\ref{eigen2}) can be obtained  performing a three-dimensional 
reduction of a Bethe-Salpeter equation with an instantaneous interaction
and  projecting the result onto the positive energy spinors~\cite{itzy}.

\section{Solution method and fitting procedure} \label{sec:determspectr}

\subsection{Solution method}
In order to determine the resonance energies of  the spectrum we have to solve Eq.~(\ref{eigen2}) 
which is a  Fredholm integral equation of second order~\cite{Arfken}.
There are different  available methods to solve this kind of integral equations.
In particular, we employ a variational procedure.
First, we express the wave function of the system by means of
a superposition of the wave functions of a given trial basis. 
Using the standard $L S$ coupling for the angular momenta the wave functions read
\begin{eqnarray}\label{basis}
\Psi_{n,\{\nu\}}(\vec{p})=R_{n,L}(p;\bar{p}) \left[ Y_L(\hat{p})\otimes\chi_S\right]_{J,M_J} ,
\end{eqnarray}
where the radial function in the momentum space $R_{n,L}(p;\bar{p})$ represents the trial function, 
$n$ is the principal quantum number,
$\bar p$ is the variational parameter with the dimension of momentum, 
$Y_{L,M_L}(\hat{p})$ is the corresponding spherical harmonic, and $\chi_{S,M_S}$ is the spin function. 
For convenience we introduce the  shorthand notation $ \{\nu\}= L,S,J$ 
for the spin-angle  quantum numbers of the basis,
i.e., the orbital angular momentum ($L$), the total spin ($S=0,1$) and 
the total angular momentum ($J$).
For  the actual  calculations we take the projection of the 
total angular momentum on the $z$-axis as $M_J=J$.
The matrix elements involving other values of $M_J$ are related to the $M_J=J$ 
matrix element through the Wigner-Eckart theorem~\cite{edmonds}.
For simplicity we do not consider the possibility that different values of $L$ mix,
whose effect has been shown to be negligible in semirelativistic models~\cite{Radford,DeSanctis2}.
For the radial part we select the orthonormal three dimensional harmonic oscillator wave functions 
\begin{equation}
R_{n,L}(p;\bar{p})= \left[ \frac{2n!} {\Gamma(n+L+\frac{3}{2})}\right]^{\frac{1}{2}}
\frac{s^L}{\bar{p}^{\frac{3}{2}}}  {\cal L}^{L+\frac{1}{2}}_n(s^2)\exp{\left[\frac{-s^2}{2}\right]}, \label{oscilador}
\end{equation}
where ${\cal L}^{L+\frac{1}{2}}_n(s^2) $ are the generalized Laguerre polynomials, 
$s \equiv p/\bar p$ and $\bar{p}$ is the variational parameter.
Now, we can diagonalize the Hamiltonian (mass) matrix in Eq.~(\ref{eigen2}) obtaining
\begin{widetext}
\begin{equation}
M_{\{\nu\},n_b,n_a}= M_0 \delta_{n_b,n_a}+
\int  d^3 p \: \Psi_{n_b, \{\nu\}}^{\dag}(\vec{p}) K(\vec{p}) \Psi_{n_a,\{\nu\}}(\vec{p})
+\int d^3 p_b \: d^3 p_a \Psi_{n_b,\{\nu\}}^{\dag}(\vec{p_b}) 
{\cal H}_\text{int} (\vec{p}_b,\vec{p}_a) \Psi_{n_a,\{\nu\}}(\vec{p_a}) ,
\label{eigen3}
\end{equation}
\end{widetext}
where, as before, the Pauli spin matrices in ${\cal H}_\text{int} (\vec{p}_b,\vec{p}_a)$ have not been explicitly written.
Notice that a six-dimensional integration over $\vec p_b$ and $\vec p_a$ appears in the interaction term.
We integrate analytically the spin-angle part of the matrix elements (see details in Appendix \ref{appendix})
so that only a three-dimensional integration over $p_b$, $p_a$ and $\cos \theta =x$ is left.

To solve Eq.~(\ref{eigen3}) we use the variational diagonalization procedure introduced in~\cite{DeSanctis2}.
We obtain good numerical convergence for the energy eigenvalues
taking the first ten trial wave functions of the basis for each state.
Hence, the $10 \times 10 $ Hamiltonian matrix is diagonalized and minimized 
through the  standard variational approach. 
Once a numerical solution is obtained, we
check that the solution represents a good numerical approximation. 
We substitute the numerical solution in the lhs of Eq.~(\ref{eigen2}) 
and verify that it fulfills the integral equation.

\subsection{Fitting and uncertainties calculation} \label{sec:fits}
The two models under study defined by potentials I and II,
see Eq.~(\ref{pot_options}), depend on seven and eight parameters respectively.
Potential I depends on 
$m$ (constituent mass of the $c$ and $\bar{c}$ quarks), 
$M_0$ (phenomenological zero point energy of the spectrum),
$\alpha_{st}$ (effective strong coupling constant),
$\beta_v$ (vector confinement intensity),
$A$ (constant scalar interaction), and
$k_{s}$ and $p_{s}$  (screening parameters),
while the model with potential II depends on the same parameters plus $\beta_s$ 
that represents the scalar confinement intensity. 

To determine the values of the parameters, the corresponding uncertainties 
and the charmonium spectrum  
we fit the experimental resonance masses given in Table~\ref{tab:table2}, 
i.e. the ground states of each $J^{PC}$ band plus the $\eta_c'$ and $\psi'$ resonances,
whose energies are below the open charm threshold $D\bar{D}$.
In doing so, we proceed as follows~\cite{Fernandez-Ramirez:2015fbq}:
\begin{enumerate}
\item We randomly choose values for the masses of the resonances by sampling 
a Gaussian distribution according to the uncertainties given in Table~\ref{tab:table2},
obtaining a resampled lowest-lying charmonia spectrum;
\item We use the least-squares method to minimize the
squared distance
\begin{equation} 
d^2 = \sum_{i} \left( E_i - M_i \right)^2,
\end{equation} 
where $M_i$ represent these resampled ground state experimental masses, 
i.e. the lowest-lying states with quantum numbers: 
$i\equiv J^{PC}=$ $0^{-+}$ ($\eta_c$ and $\eta'_c$), $1^{--}$ ($J/\psi$ and $\psi'$), 
$0^{++}$ ($\chi_{c0}$), $1^{+-}$ ($h_c$), $1^{++}$ ($\chi_{c1}$) and $2^{++}$ ($\chi_{c2}$),
and $E_i$  are the theoretical CM energies 
of the lowest-lying states obtained from the potential models I and II
solving Eq.~(\ref{eigen3}).
The fits are performed using {\sc MINUIT}~\cite{MINUIT}.
\end{enumerate}

\begin{table}
\caption{Fitted charmonia for potentials I and II compared  to the experimental masses from  PDG.
$n$ stands for the principal quantum number,
$L$ for the orbital angular momentum, 
$J$ for the total angular momentum 
and $S$ for the spin.} 
\label{tab:table2}
\begin{ruledtabular}
\begin{tabular}{c|ccccc}
Name &$n\:  ^{2S+1}L_{J}$& \multicolumn{3}{c}{Masses (MeV)} \\ 
& & Potential I  & Potential II& Experiment\\ 
\hline
$\eta_c$ &$1\: ^1S_0$&$\phantom{0}2990^{+6}_{-12}$& $\phantom{0}2981^{+  8}_{-12}$ &$\phantom{0}2983.6 \pm 0.6$  \\ 
$J/\psi$ &$1\: ^3S_1$&$3089.3^{+5.7}_{-6.5}$ & $\phantom{0}3101^{+ 13}_{-10}$ &$\phantom{0}3096.916\pm 0.011$\\
$\chi_{c0}$ &$1\: ^3P_0$&$3417.0^{+8.4}_{-5.9}$ & $\phantom{0}3416^{+ 11}_{-13}$     &$\phantom{0}3414.75\pm0.31$\\
$\chi_{c1}$ &$1\: ^3P_1$&$3500.7^{+4.4}_{-3.4}$  &  $3506.3^{+  7.7}_{-7.1}$    &$\phantom{0}3510.66\pm0.07$\\
$h_c$ &$1\: ^1P_1$&$3514.9^{+ 5.9}_{-4.6}$ & $3521.0^{+6.5}_{-6.7}$  &$\phantom{0}3525.38\pm 0.11$ \\
$\chi_{c2}$ &$1\: ^3P_2$&$\phantom{0}3579^{+5}_{-12}$ & $\phantom{0}3563^{+8}_{-14}$    &$\phantom{0}3556.20\pm 0.09$\\
$\eta_c'$&$2\: ^1S_0$&$\phantom{0}3646^{+7}_{-12}$&$\phantom{0}3647^{+10}_{-12}$ &$\phantom{0}3639.2\pm 1.2$\\
$\psi'$ &$2\:  ^3S_1$&$3674.9^{+5.4}_{-5.7}$ &$3679.2^{+  9.2}_{-9.4}$  &$3686.109^{+0.012}_{-0.014} $\\
\end{tabular}
\end{ruledtabular}
\end{table}

This procedure is repeated $1000$ times providing enough statistics 
to compute the expected value of the parameters and the spectrum energies
as well as their uncertainties at a $1\sigma$ (68\%) confidence level (CL).
The expected values of the parameters are computed as the mean values of the $1000$ samples.
To compute the uncertainties at a $1\sigma$ CL we
take the best 68\% of the fits and we compute the differences between 
the mean value and the highest and lowest masses.
In this way our uncertainties are not symmetric.
Table~\ref{tab:table1} provides the fitted parameters for both potentials
and Tables~\ref{correlation1} and~\ref{correlation2} 
the correlation matrices of the parameters 
for potentials I and II, respectively.

\begin{table}
\caption{Fit parameters for the two potentials considered in this work. 
Error bars are reported at the 1$\sigma$ (68\%) CL 
and take into account all the correlations among parameters 
(see Section~\ref{sec:fits}).} \label{tab:table1}
\begin{ruledtabular}
\begin{tabular}{c|cc}
Parameter&Potential I&Potential II\\\hline
$m$ (MeV)  &$ 1455^{+ 30}_{- 26}$     &$1364^{+46}_{-63}$\\
$M_0$ (MeV) &$\phantom{0}108^{+ 21}_{-25}$       &$\phantom{0}270\pm45$\\
$\alpha_{st}$ &$\phantom{00}0.63^{+  0.08}_{-0.06}$ &$\phantom{0}0.544^{+ 0.027}_{-0.060}$\\
$\beta_{v}$ (GeV$^2$)&$\phantom{0}0.0134^{+0.0009}_{-0.0014}$ &$\phantom{0}0.004^{+ 0.001}_{-0.002}$\\
$k_{s}$  &$\phantom{00}59^{+11}_{-9}$ &$\phantom{0}96^{+14}_{-10}$\\
$p_{s}$ (GeV) &$\phantom{0}0.698^{+ 0.065}_{-0.067}$ &$\phantom{0}0.91^{+ 0.11}_{-0.06}$\\
$A$ (GeV$^{-2}$) &$\phantom{00}0.015\pm 0.004$  &$-0.006\pm0.002$\\
$\beta_{s}$ (GeV$^2$)&$\phantom{00}0$ (fixed) &$\phantom{0}0.013^{+ 0.004}_{-0.003}$
\end{tabular}
\end{ruledtabular}
\end{table}
%

\begin{table}
\caption{Correlation matrix for the parameters of potential I.} \label{correlation1}
\begin{ruledtabular}
\begin{tabular}{c|ccccccc}
&$m$&$M_0$&$\alpha_{st}$&$\beta_v$&$k_{s}$&$p_{s}$&$A$ \\\hline
$m$ &  $\phantom{-}1$ & & &  & &\\
$M_0$ &  $-0.78$ & $\phantom{-}1$ & && &\\
$\alpha_{st}$& $\phantom{-}0.86$& $-0.40$&  $\phantom{-}1$  && &\\
$\beta_v$& $-0.88$ & $\phantom{-}0.39$ &$-0.93$ &$\phantom{-}1$& & &\\
$k_{s}$ & $\phantom{-}0.11$ &$-0.02$ & $\phantom{-}0.13$ &$-0.23$ &$\phantom{-}1$  &  &\\
$p_{s}$ & $\phantom{-}0.33$  &$-0.34$ & $\phantom{-}0.05$ &$-0.16$ &$-0.62$ & $\phantom{-}1$   & \\ 
$A$ & $-0.76$  &$\phantom{-}0.59$  &$-0.53$ &$\phantom{-}0.62$ &$\phantom{-}0.03$  &$-0.70$&$\phantom{-}1$ \\
\end{tabular}
\end{ruledtabular}
\end{table}
%

\begin{table}
\caption{Correlation matrix for the parameters of potential II.} \label{correlation2}
\begin{ruledtabular}
\begin{tabular}{c|cccccccc}
&$m$&$M_0$&$\alpha_{st}$&$\beta_v$&$k_{s}$&$p_{s}$&$A$ &$\beta_s$\\ \hline
$m$  &$\phantom{-}1$ & & && && &\\
$M_0$ &$-0.61$  &  $\phantom{-}1$ & & && & &\\
$\alpha_{st}$&$\phantom{-}0.40$ &$\phantom{-}0.31$ &$\phantom{-}1$ & & & & &\\
$\beta_v$&$-0.12$  &$\phantom{-}0.04$   &$\phantom{-}0.06$     &$\phantom{-}1$ & & & &\\
$k_{s}$ & $\phantom{-}0.31$  &$-0.10$  &$\phantom{-}0.14$ &$-0.23$ &$\phantom{-}1$ & & &\\
$p_{s}$ &$\phantom{-}0.02$  &$-0.20$  &$-0.42$   &$-0.17$ &$-0.35$ &$\phantom{-}1$ & &\\
$A$ & $\phantom{-}0.22$  &$-0.18$  &$\phantom{-}0.09$ &$-0.21$ &$\phantom{-}0.24$&$\phantom{-}0.58$ &$\phantom{-}1$ &\\
$\beta_s$& $-0.29$ &$-0.20$  &$-0.52$ &$-0.66$&$\phantom{-}0.02$ &$\phantom{-}0.33$&$\phantom{-}0.29$ &$\phantom{-}1$\\
\end{tabular}
\end{ruledtabular}
\end{table}

Table \ref{correlation1} (potential I), 
shows a strong correlation among the parameters of the vector interaction in Eq.~(\ref{vector}). 
Furthermore,  we note a marked correlation between $A$ and all the parameters of the vector interaction,  
confirming that, in this model, vector and scalar interactions are physically correlated.
Besides, it is worth highlighting  
the correlation of the parameters in the vector and scalar interactions 
with the screening parameters $k_s$ and $p_s$ of Eq.~(\ref{26}).
We notice that the parameter 
$p_s$ turns out to be  weakly correlated  
with the parameters of the vector interaction 
but strongly correlated with the parameters of the scalar interaction.  
For potential II, see Eq.~(\ref{pot_options}), 
we have the additional parameter $\beta_s$.
In this case the parameters are less correlated 
among them than for potential I, see Table~\ref{correlation2}. 
In particular, we point out that 
the correlations between the 
effective quark mass ($m$) and other parameters 
are significantly reduced in potential II compared
to potential I, with the exception of 
the phenomenological zero point energy $M_0$. 
Also, the correlation between 
$\alpha_{st}$ and $\beta_v$  is largely reduced. 
In this way, the parameters in the vector interaction are 
uncorrelated among them. 
Regarding the scalar interaction, 
we observe that the parameters $A$ and $\beta_s$ 
are almost uncorrelated.
However, we noticed a considerable 
correlation between parameters $\beta_v$ and $\beta_s$,
suggesting that, in any case, 
a strong physical correlation exists between the confinement terms 
of the vector and the scalar interactions.  
The parameter $p_s$ of the screening factors is weakly correlated 
with the other parameters of the interactions except with 
the phenomenological parameter $A$  in the scalar interaction.  
The high correlation highlights 
how the screening impacts the scalar interaction.
Finally, comparing the correlation matrices
for both potentials,  
we find that the addition 
of the confining term in the scalar interaction 
reduces the correlation among some of the parameters,
which allows for a more straightforward interpretation of the
physics associated to each term in the potential.

\begin{figure}
\begin{center}
\rotatebox{0}{\scalebox{0.3}[0.3]{\includegraphics{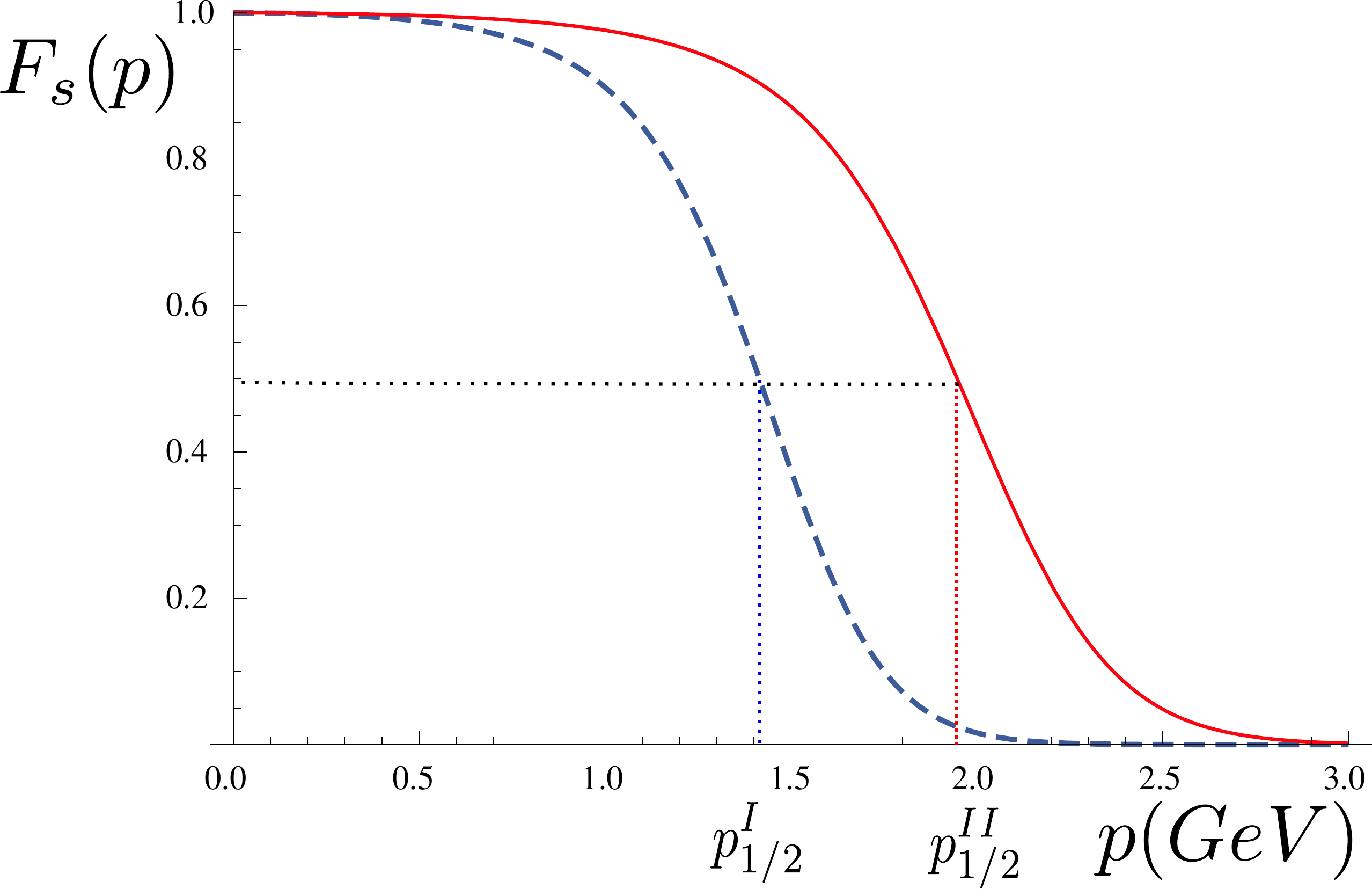}}} 
\caption{Screening function, Eq.~(\ref{26}), 
for potentials I (dashed blue) and II (solid red).
For $k_s$ and $p_s$  we use the central values in 
Table~\ref{tab:table1}.
We highlight the values of the screening momenta 
$p^\text{I}_{1/2}$ (potential I) and $p^\text{II}_{1/2}$ (potential II)
defined in Section~\ref{sec:fits}.}\label{Screening}
\end{center}
\end{figure}

For completeness, we plot the screening function 
$F_s(p)$ in Fig.~\ref{Screening} 
for the two potentials.
We introduce the screening momentum 
$p^{\: \text{j}}_{1/2}$ ($\text{j}=\text{I}, \text{II}$ 
labels potentials I and II) 
that are defined by 
$F_s(p^{\: \text{j}}_{1/2})=1/2$ (recall that $F_s(0)=1$).
We find the  values of 
$p^\text{I}_{1/2}= 1.4~\text{GeV}$ and  $p^\text{II}_{1/2}= 1.9~\text{GeV}$.
These values correspond to the screening kinetic energy
\begin{equation}
\bar{E}^{\: \text{j}}=2\sqrt{m^2 + \left( p^{\: \text{j}}_{1/2}\right) ^{\: 2} }
\end{equation}
obtaining $\bar{E}^\text{I}=4.1~\text{GeV}$ and $\bar{E}^\text{II}=4.7~\text{GeV}$.
It shows that the screening effect is active above the open charm threshold
where a sort of saturation energy must be introduced for the interaction~\cite{Xiang}.

We now consider the specific calculation of the charmonium spectrum 
and its theoretical uncertainties.
Once that $1000$ fits have been performed, 
the uncertainties of the model parameters
can be propagated to the charmonium spectrum 
using bootstrap~\cite{NumericalRecipes,Landay:2016cjw}.
This method allows us to carry to the spectrum computation 
the experimental uncertainties 
together with the correlations among the parameters.
This method is more accurate  
than those using the covariance matrix, 
although it is  computationally more expensive.
In more detail, 
the computation of the charmonium spectrum is performed as follows:
For each one of the $1000$ sets of parameters obtained from the fits we compute the spectrum.
Hence, for each state of the spectrum we have a set of $1000$ values of the mass.
The expected value of each state as well as the uncertainty 
is computed in the same way it was done for the parameters. 

\section{Results for the charmonium spectrum}\label{sec:charm}
\begin{figure*}
\begin{center}
\rotatebox{0}{\scalebox{0.55}[0.55]{\includegraphics{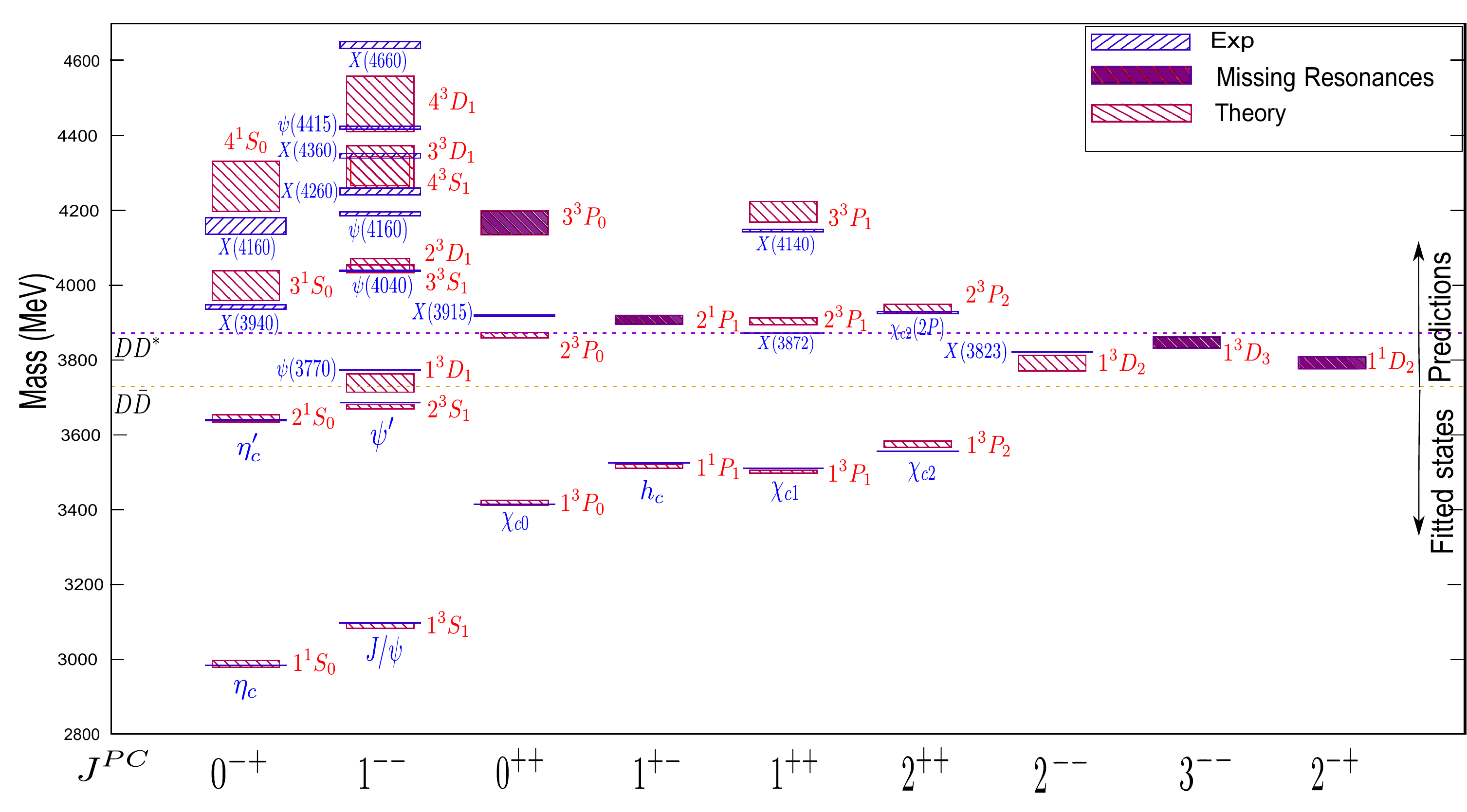}}} 
\caption{
Complete charmonium spectrum obtained with potential I (red boxes) 
compared to the experimental spectrum from  PDG (blue boxes).
The size of the red boxes accounts for error bars
at  1$\sigma$ (68\%) CL. In the same way,  
the size of the blue boxes stands for the experimental error of the resonance masses. 
The dashed lines stand for the open charm energies $D\bar{D}$ and $DD^*$. 
The calculation of the uncertainties takes 
into account all the correlations among the parameters (see Section \ref{sec:fits} for details).
In the horizontal axis the quantum numbers $J^{PC}$  are reported for the states above. 
In the right side of the figure, the arrows indicate the states (below the $D\bar{D}$ threshold)
that are used in the fitting procedure  
and the predictions of the model with potential I. 
It is  important to note that we have used the latest nomenclature by  the PDG
where the vector mesons are also named as $X$ instead of $Y$ as  
it is commonplace in the charmonia research field.
The states in both naming schemes relate as follows:
$Y(4260)\leftrightarrow X(4260)$, $Y(4360)\leftrightarrow X(4360)$, 
$Y(4660)\leftrightarrow X(4660)$,  $Y(4140)\leftrightarrow X(4140)$ 
and $Y(3915)\leftrightarrow X(3915)$.}
\label{fig:espec1}
\end{center}
\end{figure*}

\begin{figure*}
\begin{center}
\rotatebox{0}{\scalebox{0.55}[0.55]{\includegraphics{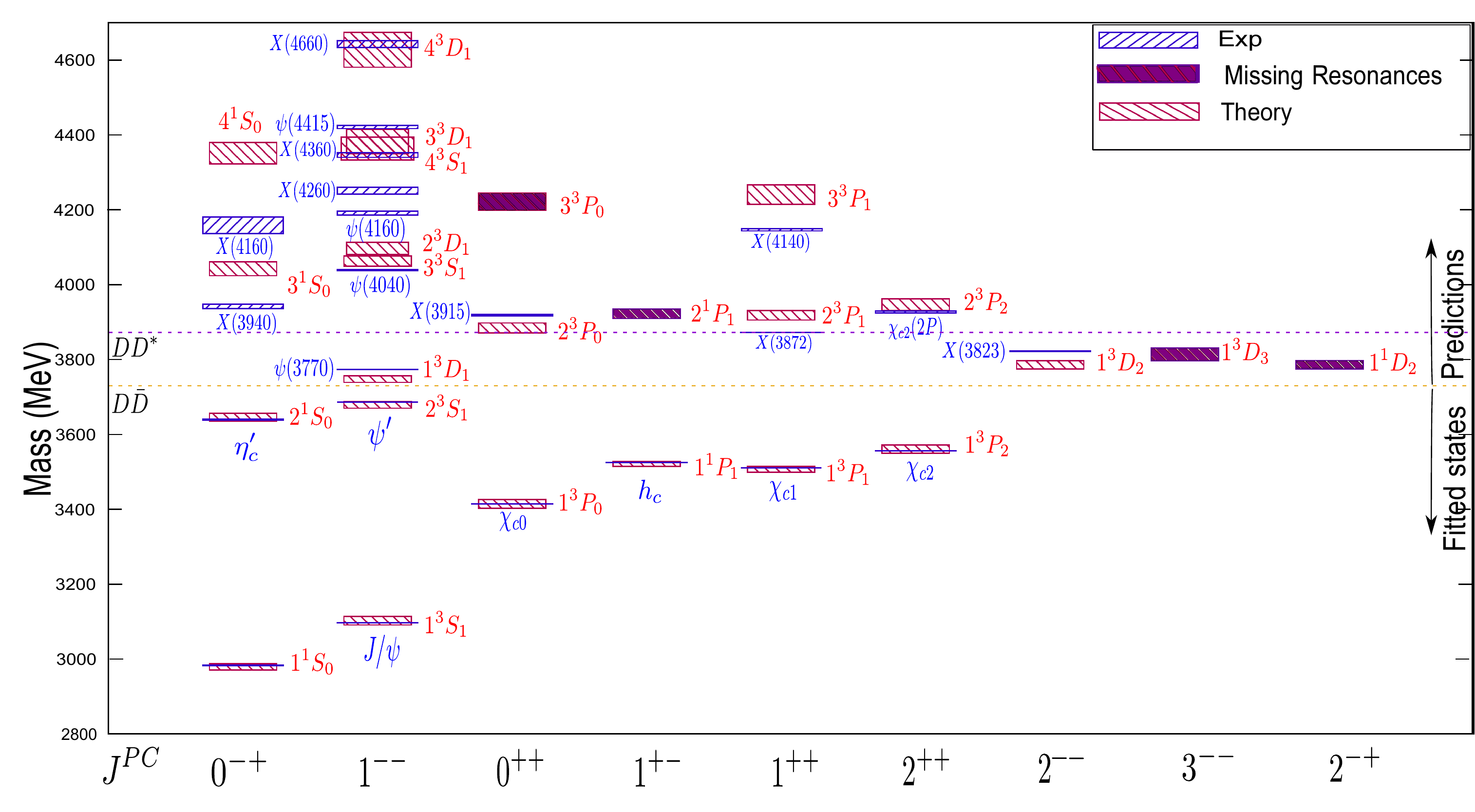}}} 
\caption{Charmonium spectrum obtained with potential II. 
Notation as in Fig.~\ref{fig:espec1}.}\label{fig:espec2}
\end{center}
\end{figure*}

Tables \ref{tab:table2} (fitted states) and \ref{tab:table3} 
(predicted states) report the obtained spectra for both models.
Figures \ref{fig:espec1} (potential I) and~\ref{fig:espec2} (potential II)
provide a graphical representation of our results compared 
to the available experimental data from PDG~\cite{PDG2016}.
We can see that the  obtained fits (below open charm threshold) 
are excellent and that, in general, 
the whole spectrum is well described with both potentials.
We note that the statistical uncertainties grow up with the
excitation energy of the states. 

%

%

\begin{table*}
\caption{
Predicted charmonia 
for  potentials I and II compared to the experimental masses and uncertainties from  PDG. 
Notation as in Table~\ref{tab:table2}. 
For the $X(4160)$  experimental mass the first error is statistical 
and the second systematic~\cite{Abe:2007sya}.}\label{tab:table3}
 \begin{ruledtabular}
\begin{tabular}{c|cccc}
Name &$n\:  ^{2S+1}L_{J}$& \multicolumn{3}{c}{Mass (MeV)} \\ 
& & Potential I  & Potential II& Experiment\\ 
\hline
---       & $1\: ^1D_2$ &$\phantom{0}3788^{+20}_{-12}$&$3789^{+9}_{-13}$& ---\\
$\psi(3770)$&$1\:  ^3D_1$&$\phantom{0}3723^{+41}_{-8}$ &$3749^{+7}_{-10}$  &$3773.13\pm 0.35$\\
$\psi(3823)$ &$1\:  ^3D_2$&$\phantom{0}3782^{+30}_{-11}$&$3788^{+9}_{-12}$&$3822.2\pm1.2$\\  
---  & $1\: ^3D_3$ &$\phantom{0}3849^{+13}_{-17}$&$3817^{+14}_{-20}$& ---\\  
$X(3915)$ [$Y(3915)$]     &$2\:  ^3P_0$&$3867.9^{+5.5}_{-8.6}$ &$3884^{+ 14}_{- 13}$ &$3918.4\pm 1.9$\\
---  & $2\: ^1P_1$ &$\phantom{0}3908^{+11}_{-13}$&$3923^{+11}_{-13}$&---\\ 
$X(3872)$ &$2\:  ^3P_1$&$3903.5^{+9.3}_{-9.4}$&$3918^{+ 13}_{- 12}$  &$3871.69\pm 0.17$ \\
$\chi_{c2}(2P)$&$2\:  ^3P_2$&$\phantom{0}3941^{+8}_{-16}$ &$3945^{+17}_{-14}$   &$3927.2\pm 2.6$\\
$\psi(4160)$ &$2\: ^3D_1$&$\phantom{0}4059^{+13}_{-20}$ &$4094^{+19}_{-20} $  &$4191\pm 5$\\
$X(3940)$ &$3\:  ^1S_0$&$\phantom{0}4030^{+9}_{-71}$ &$4043^{+ 18}_{- 19}$	   &$3942\pm 6$\\
$\psi(4040)$ &$3\:  ^3S_1$&$\phantom{0}4046^{+9}_{-13}$ &$4060^{+17}_{-11}$ &$4039\pm 1$ \\
--- &$3\:  ^3P_0$&$\phantom{0}4182^{+17}_{-47}$&$4219^{+26}_{-21}$ &---\\
$X$(4140) [$Y$(4140)]&$3\: ^3P_1$&$\phantom{0}4200^{+24}_{-31}$&$4238^{+29}_{-22}$&$4146.9\pm 3.1$\\
$X(4160)$ &$4\:  ^1S_0$&$\phantom{00}4300^{+30}_{-100}$&$4349^{+31}_{-26}$ &$4156^{+25}_{-20}\pm15$\\
\end{tabular}
\end{ruledtabular}
\end{table*}
%

\begin{table}
\caption{Predicted states that can be interpreted differently depending on the potential model.
Notation as in Table~\ref{tab:table2}.} \label{tab:tabla4}
\begin{ruledtabular}
\begin{tabular}{c|cccc}
Name & Potential&$n\:  ^{2S+1}L_{J}$& \multicolumn{2}{c}{Mass (MeV)} \\ 
&  &  & Theory & Experiment \tabularnewline
\hline 
$X(4260)$ & I & $4^{3}S_{1}$ & $4311_{-44}^{+33}$ & \multirow{2}{*}{$4251\pm9$} \tabularnewline
$[Y(4260)]$ &  II & ----- &  & \tabularnewline
\hline 
$X(4360)$&  I & $3^{3}D_{1}$ & $4312_{-53}^{+62}$ & \multirow{2}{*}{$4346\pm6$}\tabularnewline
 $[Y(4360)]$ & II & $4^{3}S_{1}$ & $4360_{-28}^{+34}$ & \tabularnewline
\hline 
\multirow{2}{*}{$\psi(4415)$}& I & $4^{3}D_{1}$ & $4489_{-78}^{+65}$ & \multirow{2}{*}{$4421\pm4$}\tabularnewline
 &  II & $3^{3}D_{1}$ & $4378_{-29}^{+36}$ & \tabularnewline
\hline 
$X(4660)$& I & ------ &  & \multirow{2}{*}{$4643\pm9$}\tabularnewline
$[Y(4660)]$ & II & $4^{3}D_{1}$ & $4620_{-39}^{+54}$ &
\end{tabular}
\end{ruledtabular}
\end{table}

Table \ref{tab:table2} shows that potential II reproduces all the 
fitted states within the $1\sigma$ CL while for potential I
$\chi_{c1}$ and $\chi_{c2}$ fall marginally outside of the $1\sigma$ CL.
Hence, potential II produces a slightly better result for the $J^{PC}$ ground states.
If we look into the predicted excited states (Table~\ref{tab:table3})
we see larger differences between both potentials,
although the produced overall structure of the spectrum is the same.

In this Section we discuss in more detail some of the states as given by potentials I and II, 
according to the $n^{2S+1}L_J$ spectroscopical assignment of the quantum numbers.

\subsection{$X(3915)$, $X(3872)$ and $\chi_{c2}(2P)$}
For these states we assign the quantum numbers $2^3 P_J$, with $J=0$, $1$ and $2$ 
for $X(3915)$, $X(3872)$ \cite{delAmoSanchez:2010jr}  and $\chi_{c2}(2P)$ respectively.
Potentials I and II are able to reproduce the hyperfine splitting of  the fitted $1P$ states 
($\chi_{c0}$, $h_c$, $\chi_{c1}$ and $\chi_{c2}$) 
but the structure of the $2P$ states cannot be reproduced with the same accuracy.
In particular, the fact that the $X(3915)$ state has a higher mass
than the $X(3872)$. However, we accurately predict the $\chi_{c2}(2P)$ 
state within uncertainties with both models.
These states lay very close to the $\bar{D}D^*$ threshold, hence, dynamical effects 
and degrees of freedom not considered in our models 
such as threshold effects and molecular admixture in the wave functions 
can be responsible for the deviations from our predictions.

The  $X(3872)$ state is of particular interest and requires a careful discussion.
This resonance was discovered by the Belle collaboration in $B$ decays~\cite{Choi} 
and was confirmed by CDF collaboration~\cite{Acosta}.
Various interpretations have been proposed for this resonance 
whose properties cannot be easily explained 
within a standard $q\bar{q}$ picture of mesons
due to the proximity to the $\bar{D} D^*$ decay threshold~\cite{Santo2}.
It has been identified as a tetraquark~\cite{Terasaki,Faustov,Esposito:2014rxa,Maiani:2004vq},
as a $\bar{D} D^*$ molecule~\cite{Wong,Baru,Ortega:2010qq,AlFiky:2005jd}, 
as $c\bar{c}$ core plus molecular components~\cite{Santo2}, 
and as a hybrid state~\cite{Takizawa}.
Even in the context of semirelativistic quark models there is 
certain controversy on the assignment of the quantum numbers.
In particular, besides the usual $2^3P_1$ assignment
(favored by studies of the decay processes~\cite{Faccini:2012zv,Aaij13}),
a $1^1D_2$ has also been proposed~\cite{DeSanctis,Radford}.
Within the Bethe-Salpeter approach this resonance 
has been identified with a $3^3P_1$ state at a mass value
of $3912~\text{MeV}$~\cite{Fischer} but, at the same time, 
a spurious $2^3P_1$ state at $3672~\text{MeV}$ is also obtained.  
Other calculations with a similar formalism obtain 
approximately the mass value of  $3900~\text{MeV}$,
but incorporating the  $X(3872)$ state in the fitting procedure~\cite{Hilger:2014nma}.
The semirelativistic calculations in~\cite{DeSanctis,DeSanctis2},
that employ a vector and a scalar interaction,
obtain a mass of $3912~\text{MeV}$ for the $2^3P_1$ state.
The UQM obtains a mass value around 
$3908~\text{MeV}$~\cite{ref,Santo2,Ferreti_alone,Ferre-Galata}. 
A calculation in the coordinate space with a screened potential 
has also given a good result of $3901~\text{MeV}$~\cite{Bai}.
In general, the nonrelativistic and semirelativistic 
quark models tend to overestimate 
the mass of this resonance with respect to the 
experimental value~\cite{Radford, Cao2012,Barnes}. 
In the present work we obtain $3903$ and $3918~\text{MeV}$ with potentials I and II, respectively. 
These results are in reasonable but not complete agreement with the experimental value.
This result together with the large isospin violation \cite{Olsen1}
which prevents the identification of this state with a standard charmonium,
indicates the need for a more accurate model that includes, 
in more precisely way, the threshold effects due to the opening of $D\bar{D}$ and $\bar{D}D^*$ channels. 
Finally, we note that both models predict a $1^1D_2$ state with mass value of $\approx3790~\text{MeV}$.
This resonance has not been experimentally observed yet.
To complete the analysis of this multiplet, we comment on the $X(3915)$ resonance. 
This state has been described as a $c\bar{c}$ state~\cite{Gang} and also has been suggested 
to be a tetraquark~\cite{Voloshin} and a molecule~\cite{Molina:2012bz}. 
In the $c\bar{c}$ context, this resonance is identified with the $2^3P_0$ state~\cite{Gang}. 
However, there exist some issues that question the validity of this assignment.
For example, it has been proposed that the $X(3915)$
is basically the $\chi_{c2}(2P)$~\cite{Zhou:2015uva,Baru:2017fgv}
Among them we highlight the lack of signal in the 
$X(3915)\rightarrow D\bar{D}$ decay channel~\cite{Olsen1}.  
Furthermore, the energy difference between the mass of the $X(3915)$
and its hyperfine splitting partner $\chi_{c2}(2P)$, 
is smaller than expected~\cite{Olsen1,Olsen2,Xia}. 
In our model the expected values of the mass for the $2^3P_0$ state are 
$\approx45~\text{MeV}$ (potential I) and $\approx20~\text{MeV}$ (potential II) below the experimental data.  
Hence, our model does not describe this state with high accuracy, but taking into account the general uncertainties
in the hadronic models, an interpretation of this state as  a $c\bar{c}$ meson cannot be excluded. 

\subsection{$\psi(3770)$, $\psi(4040)$, $\psi(4160)$ and $\psi(4415)$}
In the overpopulated  zone of the $J^{PC}=1^{--}$ resonances above the $D \bar{D}$ energy threshold, 
we find some states that are reported by  the PDG as charmonia:
$\psi(3770)$, $\psi(4040)$, $\psi(4160)$ and $\psi(4415)$,
and generally interpreted as 
$1^3D_1$, $3^3S_1$, $2^3D_1$ and $4^3S_1$ states respectively~\cite{Olsen1}.

In our model, with both potentials, we  can reproduce  
the mass of the $\psi(3770)$ with acceptable accuracy.
The minor difference  between the theoretical and experimental value ($50~\text{MeV}$ 
with potential I and $20~\text{MeV}$ with potential II) 
can be related  to the proximity of the $D\bar{D}$ threshold.   

The $\psi(4040)$, with an experimental mass value of $4039~\text{MeV}$, 
is generally identified as a $3^3S_1$ state~\cite{Olsen1, Lebed}.
However, the  theoretical mass value obtained with potential models for the $3^3S_1$ state is
greater than the experimental value~\cite{Barnes,DeSanctis2,Cao2012}.
In our case we can describe this state  with sufficient accuracy thanks 
to the inclusion of the screening functions~\cite{Bai}. 
In particular, potential I provides a theoretical prediction 
that agrees with the experimental value within uncertainties.   
Potential II also produces an acceptable mass value for this state.  

The $\psi(4160)$ is interpreted as a radial excitation of the $\psi(3770)$. 
The predicted mass value is lower than the experimental one,
a situation similar to what happens in other screened potential models~\cite{Bai,Vijande}.  
If we accept that the $\psi(4160)$ can be considered as a pure $2^3D_1$ state,  
we have to conclude that our model and the screened 
potential models are unable to reproduce this state, 
supporting a more elaborated nature for this state~\cite{Barnes-Close,Asner_etal}.  

Finally, the resonance $\psi(4415)$ has been generally interpreted 
as a $4^3S_1$ state~\cite{Olsen1} 
but also as a $5^3S_1$ state in~\cite{Bai} where the masses of the $\psi$ states 
are lowered by the inclusion of the screening effects.
Within our model it may be tentatively described in two different ways depending on which
potential is used (see Table~\ref{tab:tabla4}).
Potential I suggests that we can roughly identify this resonance 
with a $4^3D_1$ state but  potential II  suggests a $3^3D_1$ state.  
These unusual tentative assignments  for $\psi(4415)$  differ from the standard ones of 
the semirelativistic potential models~\cite{Isgur,Radford}
where this resonance is well described as a $4^3S_1$ state.
We note that in our model the $4^3S_1$ state has a mass of $4311~\text{MeV}$ 
with potential I and  $4360~\text{MeV}$ with potential II, 
that are better matches for the $X(4260)$ [$Y(4260)$] and the $X(4360)$ [$Y(4360)$]  resonances, respectively.
Hence, our $4^3S_1$ state cannot be identified as the $\psi(4415)$.

\subsection{$X(4260)$, $X(4360)$ and $X(4660)$}
These resonances  with $J^{PC}= 1^{--}$ were discovered 
by BaBar~\cite{B.Aubert1} and Belle~\cite{C.Yuan,Wang1} collaborations
and confirmed by a combined analysis of the data~\cite{Z.Liu}.
There exist different interpretations of the nature of these resonances: 
hybrid charmonia~\cite{Zhu,Guo:2008yz}, tetraquarks~\cite{Wang,Galkin-Fau}, 
molecules~\cite{Close}, $c\bar{c}$ mesons~\cite{Ding}
and some authors suggest that these signals are the product of the
interference between different channels and resonances~\cite{Chen0,Chen1,Chen2}. 
In the charmonium context, these vector mesons are atypical.  
Their quantum numbers are univocally determined by their production mechanism
but some of their features, 
e.g. the absence of open charm production~\cite{Galkin-Fau}, 
do not match the $c\bar{c}$ picture~\cite{brambilla,Olsen1}.

Our two potentials produce different results for these states.
We note that, in our model, 
the predictions  in the  high energy region of the spectrum
depend strongly on the form of the scalar interaction.
Potential I predicts a  $4^3S_1$ state, i.e. $\psi(4S)$, 
that can be roughly identified with the $X(4260)$ [$Y(4260)$] resonance.
This assignment is reasonable if we compare with the bottomonium system.
In particular, we compare the experimental mass difference 
$\Upsilon(4S)-\Upsilon(3S)\approx 224~\text{MeV}$ 
with that of $\psi(4S)-\psi(3S)\approx 254~\text{MeV}$.
In~\cite{Liu-Chen} a mass of $\approx4260~\text{MeV}$ was obtained for the $\psi(4S)$ state.
Also, as in~\cite{Bai}, in our model we can match the $3^3D_1$ state 
to the $X(4360)$  resonance.
Finally, we note that potential I cannot reproduce the $X(4660)$  [$Y(4660)$] resonance.
Potential II depicts a completely different situation.
We identify the resonances $X(4360)$  [$Y(4360)$] and $X(4660)$
with the $4^3S_1$  and  $4^3D_1$ states respectively but
we cannot describe the $X(4260)$ state. 

\subsection{$X(3940)$ and $X(4160)$}
These states were observed by Belle collaboration in double charm 
production processes~\cite{Abe:2007sya,Abe,Shen}. 
Their decays  suggest the $J^{PC}=0^{-+}$ quantum numbers.
They have been described in many different ways:
hybrids~\cite{Petrov}, molecules~\cite{Molina:2009ct,Fernandez}, 
and as the second ($\eta_c''$) and the third ($\eta_c'''$) radial excitations 
of the $\eta_c$ meson in a standard  $c\bar{c}$ picture~\cite{KTChao,Estia,Ping}. 
Both semirelativistic and nonrelativistic potential models~\cite{Barnes}
obtain masses for $\eta_c''$ and $\eta_c'''$ above the experimental values 
of $X(3940)$ and $X(4160)$,  which casts doubts on these assignments. 

In~\cite{Liu-Chen} it is assumed that the energy gap between
$\psi(3S)-\psi(2S)\approx353~\text{MeV}$ resonances is similar to the  
$\eta_c(3S) -\eta_c(2S)$ gap, making it possible to estimate the energy of $\eta(3S)$. 
The authors found that $\eta(3S)\approx3992~\text{MeV}$, 
roughly close to the experimental mass of the $X(3940)$. 
This allows us to identify it as a $3^1S_0$ state. 
On the other hand, through a decay analysis in~\cite{Liu-Chen} 
it is concluded that it is not possible to identify 
the $X(4160)$ resonance with a $4^3S_0$ state.

In the screened potential model~\cite{Bai}
the correspondence between the $X(3940)$ resonance 
and the $3^1S_0$ state is slightly improved. 
They found $\eta_c(3S)=3991~\text{MeV}$, 
but the difference between the experimental mass of the $X(4160)$ 
and the theoretical
mass value of the $4^1S_0$ state ($4250~\text{MeV}$) 
is large enough to call into question this assignment. 
In our model, with potential I and the assignments $3^1S_0$ and $4^1S_0$,
we obtain the values of $4030~\text{MeV}$ and $4302~\text{MeV}$ 
for $X(3940)$ and $X(4160)$, respectively. 
Using potential II, we obtain for the same states the values 
of $4043~\text{MeV}$ and $4349~\text{MeV}$, respectively.
We see in Table~\ref{tab:table3} and in Fig.~\ref{fig:espec1} that 
with potential I we can obtain a rough description of $X(3940)$. 
Nevertheless, in general, we conclude that our model 
is not able to describe appropriately these two resonances.   

\subsection{$X(4140)$}\label{X(4140)-state}
This resonance was first observed in~\cite{Aaltonen1} and later confirmed~\cite{Aaltonen2}. 
It has been interpreted as a tetraquark~\cite{Fang,Stancu,Wang:2015pea}, 
as a molecule~\cite{Wang:2009ue,Wang:2014gwa,Xiang}, 
as a hybrid charmonium~\cite{Mahajan} and as a $c\bar{c}$ state~\cite{Gonzalez}.  
Recently, the LHCb collaboration~\cite{Aaij} reported the $1^{++}$ quantum numbers for this resonance,
which leads to a $3^3P_1$ state interpretation,  i.e. $\chi_{c1}(3P)$, within the  $c\bar{c}$ context~\cite{Chen:2016iua}.
The mass value of the $3^3P_1$ state in the relativistic potential~\cite{Barnes} 
and screened potential~\cite{Bai} models 
is above of the mass of the $X(4140)$  [$Y(4140)$] resonance, but, for the case of the latter 
the difference is just $40~\text{MeV}$, so this interpretation is possible.
In our case, the $3^3P_1$ state is obtained with 
higher values  than   the experimental data found by LHCb for both potentials.
More precisely,  potentials I and II predict this state approximately $30~\text{MeV}$ 
and  $70~\text{MeV}$ above the experimental value, respectively.

\subsection{$\psi(3823)$} 
This resonance  was observed for the first time more than 20 years ago~\cite{Antoni}.
Recently, it was observed again by Belle  collaboration~\cite{Bhardwaj} 
and confirmed  by BESIII collaboration~\cite{Ablikim}.
In agreement with semirelativistic potential models~\cite{Barnes,Radford},
our model describes reasonably well this state with both potentials,
$\approx3785~\text{MeV}$ versus an experimental value of $3822~\text{MeV}$,
with the standard assignment $1^3D_2$ ($J^{PC}=2^{--}$).
In~\cite{Bai}, it was found a mass value of $3798~\text{MeV}$, similar to ours.
In the UQM~\cite{Santo2} a mass of $3736~\text{MeV}$ was found,
significantly lower than the experimental value.

\subsection{Missing resonances}
Our model, as the majority of the $c \bar c$ models do,
predicts (with both potentials) some states that 
have not been observed in the experiments yet.   
In  particular, we note the states $1^1D_2$, $2^1P_1$, $3^3P_0$ and $1^3D_3$.  
The $1^1D_2$ state, as already discussed, does not represent 
a good candidate for the $X(3872)$ 
and constitutes a missing resonance produced by our model.
The $2^1P_1$ state belongs to an energy region
where only the charged  mesons $Z_c(3900)$ and $Z_c(4010)$  have been found.
This region is currently empty of charmonialike states,
while the radial excitations of the $h_c$ state should lie there.
The $3^3P_0$ state that was identified as $X(4140)$ in the past
becomes a missing resonance (see  Section~\ref{X(4140)-state})
if we consider the latest results from LHCb~\cite{Aaij}.
The last missing resonance, the $1^3D_3$ state, also remains undetected 
despite that its energy value is close to the open charm threshold.

\section{Summary and conclusions} \label{sec:conclusions}
We have studied the charmonium spectrum employing a $c\bar{c}$ 
relativistic Dirac potential model in momentum space with a vector 
and a scalar interaction.
We note that only by a combination of scalar and vector
confining potentials is it possible to eliminate the empirically
contraindicated spin-orbit interaction in the interquark potential.
Positive energy Dirac spinors have been used
without performing any nonrelativistic expansion.

In our model, we have employed for the vector interaction
a one gluon exchange term plus a confining term.
For the latter term we used a standard regularized linear confining interaction
transformed to momentum space.
For the scalar interaction we have considered two possibilities: 
(i) potential I, which is given by a constant term only, namely $A$,
that accounts for an effective confinement term;
and (ii) potential II, that together with the constant term $A$ it also incorporates
a standard regularized linear confining term.
We have also incorporated phenomenological screening factors
that take into account the coupling of the $c\bar{c}$ system with 
virtual  $q\bar{q}$ states.
Potential I depends on seven parameters:
the $c$ and $\bar{c}$ quark mass $m$, 
the zero point energy $M_0$, 
the strong coupling constant $\alpha_{st}$,
the strength of the confining term in the vector interaction $\beta_v$,
the constant term of the scalar interaction $A$,
and the parameters of the screening factors $k_{s}$ and $p_{s}$.
Potential II depends on eight  parameters: the same ones as potential I plus
the strength of the confining term of the  scalar interaction $\beta_s$.

Both potential models have been fitted to the experimental masses of eight resonances 
below the open charm threshold (see Table~\ref{tab:table2}):
$\eta_c$, $J/\psi$, $\chi_{c0}$, $\chi_{c1}$, $h_c$, $\chi_{c2}$, $\eta'_{c}$ and $\psi'$.
In both cases  we obtain  a good description of these states, 
including the unusual splitting between
$h_c$, $\chi_{c1}$ and $\chi_{c2}$ resonances. 
The uncertainties in the parameters have been computed using bootstrap technique.
In this way we can propagate exactly the statistical errors in the data to both the parameters
and the resonance masses taking into account all the correlations.
We predict the energies of the resonances above the 
open charm threshold and compute the associated uncertainties 
(see Tables~\ref{tab:table3} and \ref{tab:tabla4} and
Figs.~\ref{fig:espec1} and \ref{fig:espec2}).
We predict correctly the structure of the charmonium spectrum
and we obtain a reasonable agreement with 
most of the available experimental data.
The screening effect turns out to be relevant in the description of the spectrum with both potentials,
especially at energy values above the open charm threshold.
The correlation matrices of the parameters for potential I (Table~\ref{correlation1})
and II (Table~\ref{correlation2}) 
show that there is a small correlation between the screening parameters 
and the vector interaction parameters.
However, the correlation is strong with the parameters of the scalar interaction. 
Hence, we confirm that the screening factors 
in our model take phenomenologically (and only partially) 
into account the excitation of $q\bar{q}$ degrees of freedom.
As the excitation energy grows, it becomes more difficult to reproduce the data, 
indicating that the presence of new degrees of freedom that should be introduced 
in the theoretical model.
This point requires a new and more comprehensive investigation.

We have performed a full statistical error analysis, 
determining the uncertainties of the  parameters and their correlations.
We have exactly propagated those uncertainties 
and correlations to the predicted spectrum. 
As far as we know, previous error analysis within phenomenological models 
have been very limited and incomplete.
To perform the error analysis we use the bootstrap technique. 
This method is computationally taxing but provides a rigorous treatment of the statistical uncertainties.
Rigorous error estimations allow us to assess if the inclusion of 
a new effect in the phenomenological model is necessary or not.
Moreover, the study of the correlation among the parameters
that arises from the error analysis allows us to identify 
how independent are the different pieces of the model among them.
A full error analysis is mandatory to  
identify which deviations from experimental data 
can be absorbed into the statistical uncertainties of the models
and which can be related to physics beyond the $c\bar{c}$ picture, guiding future research.

In more detail, the $X(3915)$, $X(3872)$ and $\chi_{c2}(2P)$ are identified as
$2^3P_J$ states with $J=0$, $1$, and $2$, respectively. 
The $\chi_{c2}(2P)$ is accurately reproduced.
The $X(3872)$ mass is slightly overestimated 
($3904~\text{MeV}$ with potential I and $3918~\text{MeV}$ with potential II)
signaling that this state is mostly a $c\bar{c}$ state with its mass dynamically 
modified by effects not taken into account in our models,
e.g. molecular components or open channel effects. 
The $X(3915)$ is not well described, and, hence, the 
$2P$ splitting among these states is not properly accounted for.

We also recover the general structure of the $\psi$ states
above the $D\bar{D}$ threshold, namely $\psi(3770)$  ($1^3D_1$), 
$\psi(4040)$ ($3^3S_1$), $\psi(4160)$ ($2^3D_1$) 
and $\psi(4415)$ ($4^3D_1$ with potential I and $3^3D_1$ with potential II).
However, while $\psi(3770)$ and $\psi(4040)$ are sufficiently well described,
$\psi(4160)$ and $\psi(4415)$ are not.

Our two models produce different results for the 
$X(4260)$ [$Y(4260)$], $X(4360)$ [$Y(4360)$] and $X(4660)$ [$Y(4660)$] resonances.
Potential I produces a $4^3S_1$ state that can roughly be identified with the $X(4260)$ resonance.
The higher-lying $X(4660)$ resonance is reproduced within the uncertainties
of the model with potential II but not with potential I,
and the $X(4360)$  resonance is reproduced with both potentials, 
but within very large theoretical uncertainties.
We consider that we are not able to provide a good description of 
$X(3940)$ and $X(4160)$ resonances with our model,
an indication that these states might be dominated by dynamical 
effects that go beyond the $c\bar{c}$ picture.
The $X(4140)$ resonance is approximately reproduced with potential I,
suggesting a correspondence with a $c\bar{c}$ $3^3P_1$ state;
however, it is not properly reproduced by potential II despite the large uncertainties 
of the theoretical prediction.
The $\psi(3823)$ state is also well reproduced, 
in agreement with semirelativistic potential models.

Further investigation should be devoted to understand confinement in a more complete way.
We note that potential II, 
which includes a confining term 
in both the vector and the scalar interactions, 
provides slightly better results than potential I 
(which lacks the scalar confining term),
in agreement with the standard argument 
from spin-orbit reduction 
that the vector and the scalar potentials have
approximately the same form.
In addition, 
the lower correlation among the parameters in potential II,
Table~\ref{correlation2}
compared to the correlation among the parameters 
in potential I, Table~\ref{correlation1},
allows usto make a more straightforward connection to the
physics associated to each term in potential II than
in potential I, in particular the confining interaction term.

The inclusion of relativistic effects in $q\bar{q}$ potential models 
constitutes an important step forward 
in the description of charmonia, especially the higher-lying states. 
However, it is apparent that a deeper dynamical study 
is also needed. 
One should use interactions more straightforwardly related to QCD and include other
effects such as molecular admixture, tetraquark components,
as well as open channels and threshold effects 
have to be taken into account to obtain a complete description
of the experimental states, in particular 
those whose difference with the quark model prediction
cannot be accounted for by the statistical uncertainties.

\begin{acknowledgments}
The authors thank R.~Bijker for useful discussions and comments.
This paper is part of the PhD thesis of D.M., supported by 
Departamento Administrativo de Ciencia, 
Tecnología e Innovación  (Colciencias) 
by means of the loan-Scholarship Program No.6172.
D.M. thanks Instituto de Ciencias Nucleares (ICN) at
Universidad Nacional Auton\'oma de Mexico (Mexico City, Mexico) 
for the hospitality extended to him
during the first semester 
of 2016 and, in particular, for the access to the Tochtli-ICN cluster that was used to perform part of   
the numerical calculations of this work.
The  work of C.F.-R. is supported in part by PAPIIT-DGAPA (UNAM, Mexico) Grant No. IA101717,
by CONACYT (Mexico) Grant No. 251817 and 
by Red Tem\'atica CONACYT de F\'{\i}sica en Altas Energ\'{\i}as (Red FAE, Mexico).
M.D.S. gratefully thanks INFN Sezione di Roma 3 
and Department of Physics of Roma Tre University (Rome, Italy)
for the hospitality extended to him during his sabbatical year in $2015--16$
and, in particular, for the use of the computational facilities of the Cluster Roma Tre.
\end{acknowledgments}

\appendix
\section{Spin-angle matrix elements}\label{appendix}
To calculate the spin-angle matrix elements we 
need to  factorize the corresponding operators.
To this aim we  introduce the following quantities
\begin{equation}\label{eq:FR}
F(\vec{p})=\sqrt{\frac{E(\vec{p})+m}{2E(\vec{p})}},\hspace{0.5cm} R(\vec{p})=\sqrt{\frac{E(\vec{p})-m}{2E(\vec{p})}} ,
\end{equation}
and
\begin{eqnarray}\label{eq:etathetaetc}
\eta=F(\vec{p}_a)F(\vec{p}_b),\hspace{0.5cm}\lambda=R(\vec{p}_b)R(\vec{p}_a),\nonumber\\
\theta=F(\vec{p}_b)R(\vec{p}_a),\hspace{0.5cm}\phi=F(\vec{p}_a)R(\vec{p}_a).
\end{eqnarray}

As an example, we first consider the vector term of the interaction. 
Then, we generalize the calculation to the case of the vector and scalar interactions of the model.
Substituting the Dirac four-currents of Eq.~(\ref{fourcurr}) 
in Eq.~(\ref{eq4}) and using the definitions in Eqs.~(\ref{eq:FR}) and (\ref{eq:etathetaetc}), 
we can rewrite the vector Hamiltonian as follows
\begin{widetext}
\begin{equation}
\begin{split}
\langle \vec{p}_b|H^{(v)}|\vec{p}_a\rangle=& \: V^{(v)}(\vec q)\left\{ \left[1-\frac{(\Delta E)^2}{Q^2}\right] \left[\eta^2+\eta\lambda
(\vec{\sigma}_1\cdot\hat{p}_b)(\vec{\sigma}_1\cdot\hat{p}_a)+\eta\lambda(\vec{\sigma}_2\cdot\hat{p}_b)(\vec{\sigma}_2\cdot\hat{p}_a) \right. \right. \\
+& \left. \lambda^2(\vec{\sigma}_1\cdot\hat{p}_b)(\vec{\sigma}_2\cdot\hat{p}_b)(\vec{\sigma}_1\cdot\hat{p}_a)(\vec{\sigma}_2\cdot\hat{p}_a)\right]
+\left[ \theta^2(\vec{\sigma}_1\cdot\hat{p}_a)(\vec{\sigma}_2\cdot\hat{p}_a)(\vec{\sigma}_1\cdot\vec{\sigma}_2)
+\theta\phi(\vec{\sigma}_1\cdot\hat{p}_a)(\vec{\sigma}_2\cdot\hat{p}_b)(\vec{\sigma}_1\cdot\vec{\sigma}_2)\right. \\
+&\left.\left.\theta\phi(\vec{\sigma}_2\cdot\hat{p}_a)(\vec{\sigma}_1\cdot\hat{p}_b)(\vec{\sigma}_1\cdot\vec{\sigma}_2)
+\phi^2(\vec{\sigma}_1\cdot\hat{p}_b)(\vec{\sigma}_2\cdot\hat{p}_b)(\vec{\sigma}_1\cdot\vec{\sigma}_2)\right]
\left[1+\frac{(\Delta E)^2}{Q^2}\right]\right\} .
\end{split}
\end{equation} \label{interaction}
\end{widetext}

We use the notation from Section~\ref{TotHamil} and  rename
\begin{equation}\label{rename}
\langle \vec{p}_b|H^{(v)}|\vec{p}_a\rangle=\bar{{\cal H}}^{(v)}(\vec{p}_b, \vec{p}_a),
\end{equation}
where $\bar{{\cal H}}^{(v)}(\vec{p}_b, \vec{p}_a)$ is defined in Eq.~(\ref{25}).
Recalling Eq.~(\ref{eq:q32}) for $\vec q^{\: 2}$, we can express the rhs 
of Eq.~(\ref{rename}) as the product of the functions $\bar{{\cal H}}^{(v),k}(p_b,p_a,x)$
times the  operators  depending on  $\hat{p}_a,~\hat{p}_b,~\vec{\sigma}_1$ and $\vec{\sigma}_2$. 
These operators are denoted as 
$\Lambda_k(\hat{p}_a,\hat{p}_b,\vec{\sigma}_1,\vec{\sigma}_2)$. 
We obtain a sum of eight terms
\begin{equation}\label{eq9}
\bar{{\cal H}}^{(v)}(\vec{p}_b, \vec{p}_a)=\sum_{k=1}^{8} \bar{{\cal H}}^{(v),k}(p_b,p_a,x)
\Lambda_k(\hat{p}_a,\hat{p}_b,\vec{\sigma}_1,\vec{\sigma}_2).
\end{equation}

The explicit expressions of the first four terms of this sum are 
\begin{alignat}{2}
\Lambda_1(\hat{p}_a,\hat{p}_b,\vec{\sigma}_1,\vec{\sigma}_2)&
=\mathbb{1} , \\
\Lambda_2(\hat{p}_a,\hat{p}_b,\vec{\sigma}_1,\vec{\sigma}_2)&
= (\vec{\sigma}_1\cdot\hat{p}_b)(\vec{\sigma}_1\cdot\hat{p}_a), \\
\Lambda_3(\hat{p}_a,\hat{p}_b,\vec{\sigma}_1,\vec{\sigma}_2)&
=(\vec{\sigma}_2\cdot\hat{p}_b)(\vec{\sigma}_2\cdot\hat{p}_a),\\
\Lambda_4(\hat{p}_a,\hat{p}_b,\vec{\sigma}_1,\vec{\sigma}_2)&
=(\vec{\sigma}_1\cdot\hat{p}_b)(\vec{\sigma}_2\cdot\hat{p}_b)(\vec{\sigma}_1\cdot\hat{p}_a)(\vec{\sigma}_2\cdot\hat{p}_a),\\
\bar{{\cal H}}^{(v),1}(p_b,p_a,x)& = V^{(v)}(\vec q)\left[1-\frac{(\Delta E)^2}{Q^2}\right]\eta^2 ,\\
\bar{{\cal H}}^{(v),2}(p_b,p_a,x)&= V^{(v)}(\vec q)\left[1-\frac{(\Delta E)^2}{Q^2}\right]\eta\lambda,\\
\bar{{\cal H}}^{(v),3}(p_b,p_a,x)& =V^{(v)}(\vec q)\left[1-\frac{(\Delta E)^2}{Q^2}\right]\eta\lambda,\\
\bar{{\cal H}}^{(v),4}(p_b,p_a,x)& =V^{(v)}(\vec q)\left[1-\frac{(\Delta E)^2}{Q^2}\right]\lambda^2.
\end{alignat}

Analogously, the other terms of the vector interaction and those 
of the scalar interaction can be found through straightforward calculations.
Next, we perform a double spherical harmonic expansion on both the scalar and vector terms
with respect to the angles of $\hat{p}_b$ and of $\hat{p}_a$ for each  $\bar{\cal H}^{(\tau),k}(p_b,p_a,x)$ term, 
where  $\tau$ labels the  vector ($v$) and scalar ($s$) interactions.
We obtain
\begin{equation}
\begin{split}
\bar{\cal H}_\text{int}(\vec{p}_b, \vec{p}_a)=& \: 4\pi \sum_{\tau}  
\sum_{k}\sum_\ell{\bar{\cal H}}^{(\tau), k}_{\ell}(p_b,p_a) \\
\times  \sum_m &Y_{\ell,m}(\hat{p}_b)Y_{\ell,m}^*(\hat{p}_a)
\Lambda_k(\hat{p}_a,\hat{p}_b,\vec{\sigma}_1,\vec{\sigma}_2) ,
\end{split}\label{eq:spangf}
\end{equation}
where we have expanded in Legendre polynomials $P_\ell(x)$
\begin{equation}\label{eq12}
\bar{\cal H}_\ell^{(\tau),k}(p_b,p_a)=\frac{1}{2}\int_{-1}^{1}dx \: \bar{\cal H}^{(\tau),k}(p_b,p_a,x)P_\ell(x).
\end{equation}

The second line of Eq.~(\ref{eq:spangf}) represents the spin-angle operators
that can be calculated analytically from the angular part of the wave functions of our basis given in Eq.~(\ref{basis}).
In doing so, we write
\begin{widetext}
\begin{equation}
O^{k}_{\ell;L,S,J} 
=\int d \Omega_b\: d\Omega_a   \left[ Y_L(\hat{p}_b)\otimes\chi_S\right]_{J,J}^\dag
 \sum_m Y_{\ell,m}(\hat{p}_b)Y_{\ell,m}^*(\hat{p}_a)\Lambda_k(\hat{p}_a,\hat{p}_b,\vec{\sigma}_1,\vec{\sigma}_2) 
 \left[ Y_L(\hat{p}_a)\otimes\chi_S\right]_{J,J},
 \label{eq:spangme}
\end{equation}
\end{widetext}
where $d \Omega_a$ and $d\Omega_b$ represent the integration
elements over the angles of $\hat{p}_a$ and $\hat{p}_b$,
respectively.
The matrix elements of the previous equation can be calculated 
in a standard way by using angular momentum algebra.

Finally, the interaction matrix elements, that is the third term in the rhs
of Eq.~(\ref{eigen3}) can be written as
\begin{widetext}
\begin{equation}
\begin{split}
 \int d^3 p_b \: d^3 p_a\:  \Psi_{n_b,\{\nu\}}^{\dag}(\vec{p}_b) {\cal H}_\text{int} (\vec{p}_b,\vec{p}_a)   \Psi_{n_a,\{\nu\}}(\vec{p}_a)&  \\ 
 = 4 \pi  \sum_{\tau}  \sum_{k}\sum_\ell   O^{k}_{\ell;L,S,J}   \int_0^\infty  p_b^2 \: dp_b  \int_0^\infty p_a^2 \: dp_a \:   &
 R_{n_b,L}(p_b;\bar{p}) F_s(p_b) \bar{\cal H}_\ell^{(\tau),k}(p_b,p_a)  F_s(p_a)R_{n_a,L}(p_a;\bar{p}) ,
 \end{split}
 \end{equation}
\end{widetext}
where two integrations over $p_a$, $ p_b$
and one integration over $x=\cos \theta$ from Eq.~(\ref{eq12}) remain.


\end{document}